%% file: bmc_article.tex
\documentclass{bmcart}

\usepackage[utf8]{inputenc} %unicode support
\usepackage{graphicx,comment,xcolor,setspace}
\usepackage{subfigure}

\usepackage{booktabs}
\usepackage[ruled,vlined]{algorithm2e}
\usepackage{multirow}
\usepackage{hyperref}
\startlocaldefs
\endlocaldefs

\begin{document}

%%% Start of article front matter
\begin{frontmatter}

\begin{fmbox}
\dochead{Research}


\title{DI2: prior-free and multi-item discretization of biomedical data and its applications}

%%%%%%%%%%%%%%%%%%%%%%%%%%%%%%%%%%%%%%%%%%%%%%
%%                                          %%
%% Enter the authors here                   %%
%%                                          %%
%% Specify information, if available,       %%
%% in the form:                             %%
%%   <key>={<id1>,<id2>}                    %%
%%   <key>=                                 %%
%% Comment or delete the keys which are     %%
%% not used. Repeat \author command as much %%
%% as required.                             %%
%%                                          %%
%%%%%%%%%%%%%%%%%%%%%%%%%%%%%%%%%%%%%%%%%%%%%%
%\author[Alexandre, Costa and Henriques]{\large Leonardo Alexandre\,$^{\text{\sfb 1,2,3,}*}$, Rafael S. Costa\,$^{\text{\sfb 1,4,}*}$ and Rui Henriques\,$^{\text{\sfb 2,3,}*}$}

\author[
   addressref={aff1,aff2,aff3},                   % id's of addresses, e.g. {aff1,aff2}
   corref={aff1,aff2,aff3,aff4},                       % id of corresponding address, if any
   noteref={},                        % id's of article notes, if any
   email={leonardoalexandre@tecnico.ulisboa.pt}   % email address
]{\inits{LA}\fnm{Leonardo} \snm{Alexandre}}

\author[
   addressref={aff1,aff4},                    % id's of article notes, if any
   email={rafael.s.costa@tecnico.ulisboa.pt}   % email address
]{\inits{RSC}\fnm{Rafael S.} \snm{Costa}}

\author[
   addressref={aff2,aff3},              % id's of article notes, if any
   email={rmch@tecnico.ulisboa.pt}   % email address
]{\inits{RH}\fnm{Rui} \snm{Henriques}}

%\author[
%   addressref={aff1},                   % id's of addresses, e.g. {aff1,aff2}
%   corref={aff1},                       % id of corresponding address, if any
%   noteref={n1},                        % id's of article notes, if any
%   email={jane.e.doe@cambridge.co.uk}   % email address
%]{\inits{JE}\fnm{Jane E} \snm{Doe}}
%\author[
%   addressref={aff1,aff2},
%   email={john.RS.Smith@cambridge.co.uk}
%]{\inits{JRS}\fnm{John RS} \snm{Smith}}

%%%%%%%%%%%%%%%%%%%%%%%%%%%%%%%%%%%%%%%%%%%%%%
%%                                          %%
%% Enter the authors' addresses here        %%
%%                                          %%
%% Repeat \address commands as much as      %%
%% required.                                %%
%%                                          %%
%%%%%%%%%%%%%%%%%%%%%%%%%%%%%%%%%%%%%%%%%%%%%%

\address[id=aff1]{%                           % unique id
  \orgname{IDMEC, Instituto Superior Técnico, Universidade de Lisboa}, % university, etc
  %\street{Waterloo Road},                     %
  %\postcode{}                                % post or zip code
  %\city{Lisbon},                              % city
  \cny{Portugal}                                    % country
}
\address[id=aff2]{%                           % unique id
  \orgname{INESC-ID}, % university, etc
  %\street{Waterloo Road},                     %
  %\postcode{}                                % post or zip code
  %\city{Lisbon},                              % city
  \cny{Portugal}                                    % country
}
\address[id=aff3]{%                           % unique id
  \orgname{Instituto Superior T\'{e}cnico, Universidade de Lisboa}, % university, etc
  %\street{Waterloo Road},                     %
  %\postcode{}                                % post or zip code
  %\city{Lisbon},                              % city
  \cny{Portugal}                                    % country
}
\address[id=aff4]{%                           % unique id
  \orgname{LAQV-REQUIMTE, DQ, Faculty of Science and Technology, NOVA University of Lisbon, Campus de Caparica}, % university, etc
  %\street{Waterloo Road},                     %
  \postcode{2829-516}                                % post or zip code
  \city{Caparica},                              % city
  \cny{Portugal}                                    % country
}
%\address[id=aff2]{%
%  \orgname{Marine Ecology Department, Institute of Marine Sciences Kiel},
%  \street{D\"{u}sternbrooker Weg 20},
%  \postcode{24105}
%  \city{Kiel},
%  \cny{Germany}
%}

%%%%%%%%%%%%%%%%%%%%%%%%%%%%%%%%%%%%%%%%%%%%%%
%%                                          %%
%% Enter short notes here                   %%
%%                                          %%
%% Short notes will be after addresses      %%
%% on first page.                           %%
%%                                          %%
%%%%%%%%%%%%%%%%%%%%%%%%%%%%%%%%%%%%%%%%%%%%%%

\begin{artnotes}
%\note{Sample of title note}     % note to the article
%\note[id=n1]{Equal contributor} % note, connected to author
\end{artnotes}

\end{fmbox}% comment this for two column layout

%%%%%%%%%%%%%%%%%%%%%%%%%%%%%%%%%%%%%%%%%%%%%%
%%                                          %%
%% The Abstract begins here                 %%
%%                                          %%
%% Please refer to the Instructions for     %%
%% authors on http://www.biomedcentral.com  %%
%% and include the section headings         %%
%% accordingly for your article type.       %%
%%                                          %%
%%%%%%%%%%%%%%%%%%%%%%%%%%%%%%%%%%%%%%%%%%%%%%

\begin{abstractbox}

\begin{abstract} % abstract
\textbf{Motivation:} A considerable number of data mining approaches for biomedical data analysis, including state-of-the-art associative models, require a form of data discretization. Although diverse discretization approaches have been proposed, they generally work under a strict set of statistical assumptions which are arguably insufficient to handle the diversity and heterogeneity of clinical and molecular variables within a given dataset. In addition, although an increasing number of symbolic approaches in bioinformatics are able to assign multiple items to values occurring near discretization boundaries for superior robustness, there are no reference principles on how to perform multi-item discretizations. \\%to handle mixed data with arbitrary  on available domain knowledge  others use outcome variables to best guide the discretization process. 
\noindent\textbf{Results:} In this study, an unsupervised %fully autonomous, non-parametric and prior-free 
discretization method, DI2, for %mixed
variables with arbitrarily skewed distributions is proposed. DI2 provides robust guarantees of generalization by placing data corrections using the Kolmogorov-Smirnov test before statistically fitting distribution candidates. DI2 further supports multi-item assignments. Results gathered from biomedical data show its relevance to improve classic discretization choices.\\ %suppo, as well An arbitrarily-high number of distributions  %continuous variables according to the best fitting continuous distribution. 
%Considering the purpose of the discretized data (classification tasks, pattern mining, machine learning), the discretization method purposed can consider border values and create intermediate values (values which belong to two categories).\\
% \\
\noindent \textbf{Software:} available at \href{https://github.com/JupitersMight/DI2}{https://github.com/JupitersMight/DI2} \\
\end{abstract}

%%%%%%%%%%%%%%%%%%%%%%%%%%%%%%%%%%%%%%%%%%%%%%
%%                                          %%
%% The keywords begin here                  %%
%%                                          %%
%% Put each keyword in separate \kwd{}.     %%
%%                                          %%
%%%%%%%%%%%%%%%%%%%%%%%%%%%%%%%%%%%%%%%%%%%%%%


% MSC classifications codes, if any
%\begin{keyword}[class=AMS]
%\kwd[Primary ]{}
%\kwd{}
%\kwd[; secondary ]{}
%\end{keyword}

\end{abstractbox}
%
%\end{fmbox}% uncomment this for twcolumn layout

\end{frontmatter}

%%%%%%%%%%%%%%%%%%%%%%%%%%%%%%%%%%%%%%%%%%%%%%
%%                                          %%
%% The Main Body begins here                %%
%%                                          %%
%% Please refer to the instructions for     %%
%% authors on:                              %%
%% http://www.biomedcentral.com/info/authors%%
%% and include the section headings         %%
%% accordingly for your article type.       %%
%%                                          %%
%% See the Results and Discussion section   %%
%% for details on how to create sub-sections%%
%%                                          %%
%% use \cite{...} to cite references        %%
%%  \cite{koon} and                         %%
%%  \cite{oreg,khar,zvai,xjon,schn,pond}    %%
%%  \nocite{smith,marg,hunn,advi,koha,mouse}%%
%%                                          %%
%%%%%%%%%%%%%%%%%%%%%%%%%%%%%%%%%%%%%%%%%%%%%%

%%%%%%%%%%%%%%%%%%%%%%%%% start of article main body
% <put your article body there>

\input{Introduction}
\input{background}

\input{discussion}

\section{Conclusion}

This work proposed a new unsupervised method for data discretization, DI2, that takes into account data distribution, outliers within the data, and border values.

Results show that DI2 is a viable and coherent form of discretizing data. When compared with other well-known and frequently used unsupervised methods, DI2 brings out the same average accuracy in well-known and supervised classfiers. If we consider border values, which other unsupervised methods do not consider, and use a classifier that handles border values, FleBiC, then DI2 is able to improve the average accuracy previously achieved without border values. 

%\nocite{oreg,schn,pond,smith,marg,hunn,advi,koha,mouse}

%%%%%%%%%%%%%%%%%%%%%%%%%%%%%%%%%%%%%%%%%%%%%%
%%                                          %%
%% Backmatter begins here                   %%
%%                                          %%
%%%%%%%%%%%%%%%%%%%%%%%%%%%%%%%%%%%%%%%%%%%%%%

\begin{backmatter}

%\section*{Data and software availability}
%    The software is available at \href{https://github.com/JupitersMight/DI2}{https://github.com/JupitersMight/DI2}.
    
%    The data is publicly available at the UCI Machine Learning repository \cite{asuncion2007uci}.

%\section*{Competing interests}
%  The authors declare that they have no competing interests.
  
%\section*{Publication consent}
%  All authors consent with the publication of the manuscript.

%\section*{Authors contributions}
%  All authors contributed to the design of the methodology. LA implemented the software and produced the first draft of the manuscript. All authors revised the manuscript.

%\section*{Funding}
\section*{Acknowlegments}
  This work was supported by Fundação para a  Ciência  e  a  Tecnologia  (FCT),  through  IDMEC,  under  LAETA  project (UIDB/50021/2020), IPOscore with reference (DSAIPA/DS/0042/2018), and  ILU  (DSAIPA/DS/0111/2018). This  work  was  further  supported  by the Associate Laboratory for Green Chemistry (LAQV), financed by national funds from FCT/MCTES (UIDB/50006/2020 and UIDP/50006/2020), INESC-ID plurianual (UIDB/50021/2020) and the contract CEECIND/01399/2017 to R. S. Costa.
  
%\section*{Acknowlegments}
%    N/A
    
%\section*{Ethics approval and consent to participate}
%  N/A
%%  After submission of the .TEX file,                     %%
%%  you will be prompted to submit your .BBL file.         %%

% if your bibliography is in bibtex format, use those commands:
\bibliographystyle{bmc-mathphys} % Style BST file (bmc-mathphys, vancouver, spbasic).
\bibliography{bmc_article}      % Bibliography file (usually '*.bib' )

\end{backmatter}
\end{document}

%% file: Introduction.tex
\section{Introduction}

Approaches to discretization of continuous variables have long been discussed alongside their pros and cons. Altman \textit{et al.} \cite{altman2014categorizing} and Bennette \textit{et al.} \cite{bennette2012against} both discuss the relevance and impact of categorizing continuous variables and reducing the cardinality of categorical variables. %explanatory variables, when creating binary splits or several groups. 
Liao \textit{et al.} \cite{liao2002appropriate} compares various categorization techniques in the context of classification tasks in medical domains, without using domain knowledge of field experts. The relevance of discretization meets both descriptive and predictive ends, encompassing state-of-the-art approaches such as pattern-based biclustering \cite{henriques2014bicpam} and associative models such as XGBoost \cite{chen2016xgboost}. 

In this work we present DI2, a Python library that extends non-parametric tests to find the best fitting distribution for a given variable and discretize it accordingly. DI2 offers three major contributions: i) corrections to the empirical distribution before statistical fitting to guarantee a more robust approximation of candidate distributions; ii) efficient statistical fitting of 100 state-of-the-art theoretical distributions; and, finally, iii) assignment of multiple items according to the proximity of values to the boundaries of discretization, a possibility supported by numerous symbolic approaches \cite{okada2007exhaustive, henriques2014bicpam, zhang2017hierarchical}. The assignment of multiple items \cite{wang2017multi,wang2018multi}, generally referred as multi-item discretization, conferes the possibility to avail the wealth of data structures and algorithms from the text processing and bioinformatics communities without the risks of the well-studied item-boundaries problem.

%% file: background.tex
\section{Background}

Discretization methods have wide taxonomy\cite{garcia2012survey} with a determinant division in: 1) supervised, where the method uses the class variable to bin the data, and, 2) unsupervised, where the method is independent of the class variable. DI2 places itself on the latter, it works independently on the class variable. Other characteristics of DI2 are: %1) parametric, the number of intervals for each variable must be given,
1) static, where discretization of the variables takes place prior to an algorithm; 2) global, uses information about the variable as a whole to make the partitions and can still be applied with a scarce number of observations; 3) direct and splitting, splits the whole the range of values into \textit{k} intervals simultaneously; and 4) multivariate and univariate, DI2 can use either the whole dataset to create the intervals and discretize each variable or use each variable individually to create the respective intervals. 

\begin{figure}[!ht]
    \centering
    \includegraphics{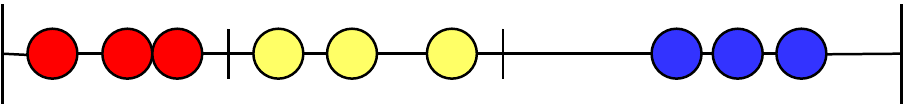}
    \caption{Illustration of equal-frequency method with 9 points along an axis and 3 categories. This method is based on the frequency of the items, where each category has the same number of items, in order to set the intervals.}
    \label{equal_freq}
\end{figure}

\begin{figure}[!ht]
    \centering
    \includegraphics{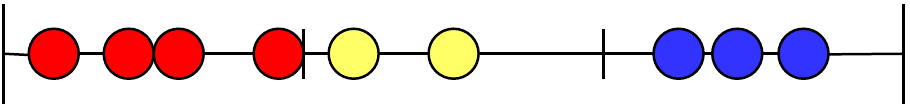}
    \caption{Illustration of equal-width method with 9 points along an axis and 3 categories. This method is based on the range taken by the items, where each category has the same width interval.}
    \label{equal_width}
\end{figure}

\begin{figure}[!ht]
    \centering
    \includegraphics[width=0.75\textwidth]{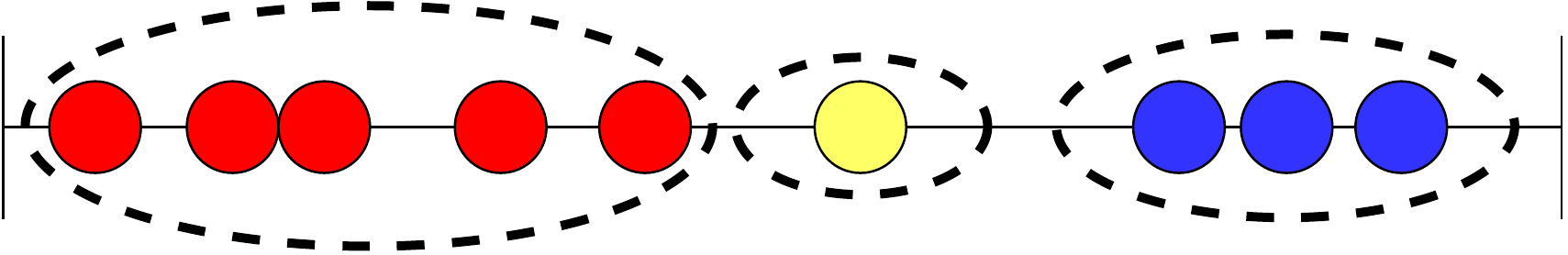}
    \caption{Illustration of K-means method with 9 points along an axis and 3 categories. This method is based in the k-means clustering, where each category is defined by a centroid.} %The method has two steps: 1) Randomly pick 1 item as the starting poing of a centroid (number of centroid is equal to the number of categories), 2) Group items by closest to each centroid and calculate new position of the centroid, this distance is based on the average distance between all points in the group. The method repeats step 2 until no centroid position is updated.}
    \label{k_means_background}
\end{figure}

Some examples of unsupervised discretization methods are PD (Proportional Discretization), FFD (Fixed Frequency Discretization)\cite{yang2009discretization}, equal-width/frequency, k-means \cite{tou1974pattern}. In this work, DI2 is compared with such classic discretization methods. These are illustrated in Figures \ref{equal_freq}, \ref{equal_width}, and \ref{k_means_background}.

\subsection{DI2: normalization and feature scaling}

%While not mandatory, data normalization and feature scaling are transformations often used either to bring out inherent properties of the data or transform the data into the same range of values. With this in mind DI2 provides normalization methods and feature scaling, which are selected for preprocessing a given variable based on its empirical distribution. The supported methods are:
While not mandatory, DI2 supports the following techniques: 1) min-max scaling,

\begin{equation}
    X'= \frac{X - X_{min}}{X_{max} - X_{min}},
\end{equation}

\noindent where $X$ is an ordered set of observed values; 2) z-score normalization, usually applied to samples normally distributed \cite{dodge2006oxford},

\begin{equation}
    X'= \frac{X - \mu}{\sigma},
\end{equation}

\noindent where $X$ is an ordered set of observed values; and 3) mean normalization,

\begin{equation}
    X'= \frac{X - \mu}{X_{max} - X_{min}}.
\end{equation}

\subsection{DI2: statistical hypotheses}

In order to discretize the data into intervals, DI2 provides two statistical hypothesis tests to be applied: 1) $\tilde{\chi}^2$ test \cite{lowry2014concepts}, and 2) Kolmogorov-Smirnov goodness-of-fit test \cite{gonzalez1977efficient}.

In the aforementioned tests the observed distribution is matched with a theoretical continuous distribution\footnote{https://docs.scipy.org/doc/scipy/reference/stats.html} provided by the SciPy open-source library \cite{virtanen2020scipy} where the parameters are estimated through maximum likelihood estimation. The binning of the distributions for the $\tilde{\chi}^2$ test is based on the number of categories the user inputs and are built using the cumulative distribution function. The user can either choose the $\tilde{\chi}^2$ or the Kolmogorov-Smirnov goodness-of-fit as the \textit{primary} fitting test. %for ranking. and selecting distributions. %and select the best statistical fitting.
The theoretical continuous distribution with the lowest test statistic is picked as the best fit for the observed distribution.

\subsection{DI2: outlier correction}

The Kolmogorov-Smirnov goodness-of-fit test can optionally be used to remove up to 5\% outlier points, from the observed distribution, according to the theoretical continuous distribution under assessment. Kolmogorov-Smirnov goodness-of-fit test returns a statistic (D statistic) that represents the maximum distance between the observed that and the theoretical continuous distribution we are testing.

\begin{equation}
    D = max\{\max_{1 \leq j \leq n} \{\frac{j}{n} - F(X_j)\}, \max_{1 \leq j \leq n} \{F(X_j) - \frac{(j-1)}{n}\},
\end{equation}
    
%\begin{equation}
   % D+ = \max_{1 \leq j \leq n} \{\frac{j}{n} - F(X_j)\},
%\end{equation}

%\begin{equation}
%    D- = \max_{1 \leq j \leq n} \{F(X_j) - \frac{(j-1)}{n}\}.
%\end{equation}

\noindent where $n$ is the number of samples, $j$ is the index of a given sample, and $F$ is the frequency of sample $X_j$. Using this statistic we can pin point where the farthest point between the distributions is and remove it. After up to 5\% of the samples have been removed, the iteration with the best Kolmogorov-Smirnov statistic is picked (from 0 outliers removed to up to 5\%). The data produced by outlier removal is then used to run the main statistical hypothesis test picked ($\tilde{\chi}^2$ or Kolmogorov-Smirnov). %Suppose we have 100 samples and we are using Outlier removal, we will execute Kolmogorov-Smirnov goodness-of-fit test we will remove up to 5 points and pick the best iteration and use the data
This correction guarantees the absence of penalizations caused by abrupt yet spurious deviations driven by the selected histogram granularity and help consolidate the choice of the theoretical continuous distribution.
%The modified observed distribution from the iteration of the Kolmogorov-Smirnov test with the best KS-statistic is used for the subsequent fitting stage. 

\subsection{DI2: multi-item discretization}

After selecting the theoretical continuous distribution that best fits the continuous variable, DI2 proceeds with the discretization. Given a desirable number of categories (bins), multiple cut-off points are generated using the inverse cumulative distribution function of the theoretical continuous distribution. The cut-off points guarantee an approximately uniform distribution of observation per category, although empirical-theoretical distribution differences can underlie imbalances. 

\begin{figure}[!ht]
    \centering
    \includegraphics[width=0.75\textwidth]{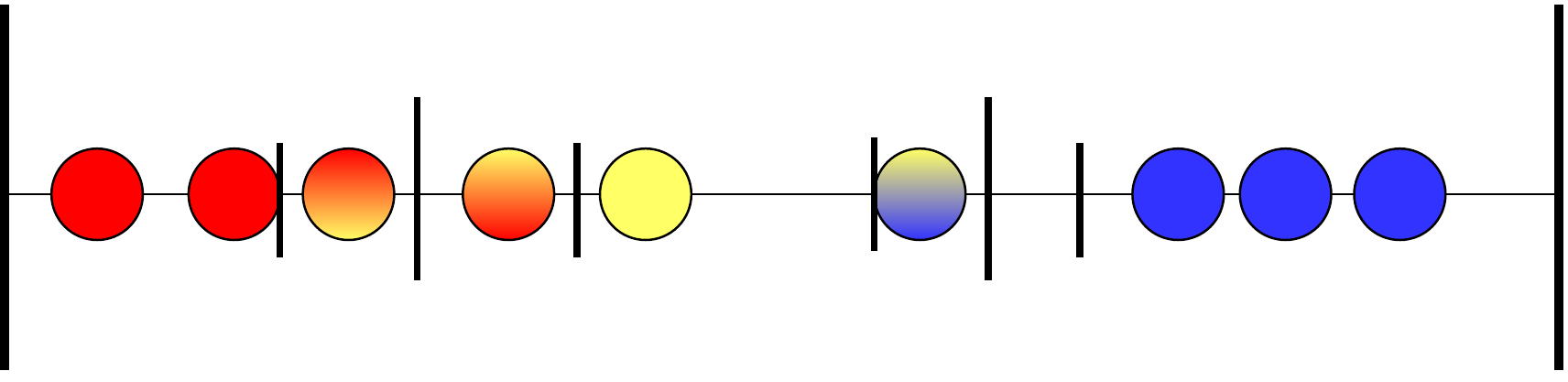}
    \caption{Illustration example of discretization with 9 points along an axis and 3 categories considering border values (values which belong to 2 categories).}
    \label{bordervalues_background}
\end{figure}

DI2 supports multi-item assignments by identifying border values for each category, this is exemplified in Figure \ref{bordervalues_background}. To this end, the user can optionally also define a percentage (between 0 and 50\% with 20\% default) to affect the width of the borders. These borders take an intermediate value which symbolize that it belongs to both upper and lower category. Width extremes, 0\% (50\%) correspond to none (one) additional category assigned to every observation. 

\section{Implementation}

DI2 tool is fully implemented in Python and is provided as an open-source method at GitHub with  well-annotated APIs and notebook tutorials for a practical illustration of its major functionalities. The algorithm workflow is shown in Algorithm 1 and the Kolmogorov-Smirnov correction is shown in Algorithm 2.

\begin{algorithm}[!ht]
\footnotesize
\SetAlgoLined
\KwIn{dataset, number\_of\_bins}
\textbf{Optional input:} statistical\_test="chi2", multi\_item\_cutoff\_margin=0.2, kolmogorov\_opt=True, normalizer="min\_max", distributions=[...], single\_column\_discretization=True

\KwOut{The dataset discretized}
 y\_normalized = []\;
 \eIf{single\_column\_discretization}
 {
    \For{column in dataset.columns}{
        y\_normalized = normalization(dataset[column],normalizer)\;
        main\_operation(...)\;
    }
 }
 {
    \For{column in dataset.columns}{
        y\_normalized.append(normalization(dataset[column],normalizer))\;
    }
    main\_operation(...)\;
 }
 \SetKwFunction{FMain}{main\_operation}
 \SetKwProg{Pn}{Function}{:}{\KwRet}
 \Pn{\FMain{...}}{
    best\_dist, test\_statistic, data\_used\;
    \For{distribution in distributions}{
        temp\_statistic\;
        \eIf{statistical\_test == "chi2"}{
            temp\_data = kolmogorov\_goodness\_of\_fit(y\_normalized, distribution, kolmogorov\_opt)\;
            temp\_statistic = chi\_squared\_goodness\_of\_fit(temp\_data, distribution, number\_of\_bins)\;
        }{
        temp\_statistic, temp\_data = kolmogorov\_goodness\_of\_fit(y\_normalized, distribution, kolmogorov\_opt)\;
        }
        \If{temp\_statistic $<$ test\_statistic}{
            test\_statistic = temp\_statistic\;
            best\_dist = distribution\;
            data\_used  = temp\_data\;
        }
    }
    dataset[column] = discretize(best\_dist, multi\_item\_cutoff\_margin, data\_used, dataset[column], number\_of\_bins, y\_normalized)\;
 
  }
 \caption{DI2 main algorithm}
\end{algorithm}

\begin{algorithm}[!ht]
\footnotesize
\SetAlgoLined
\KwIn{observed\_distribution, theoretical\_distribution, outlier\_removal\_flag}
\KwOut{The statistic of Kolmogorov test and the corresponding data}
 N5 = size(data) $\times $ 0.05 \textbf{if} outlier\_removal\_flag \textbf{else} 1\;
 results = []\;
 i = 0\;
 \While{i $<$ N5}{
    Estimate\_Parameters(theoretical\_distribution)\;
    D\_plus = D\_minus = []\;
    idx\_max\_d\_plus = idx\_max\_d\_minus = []\;
    calculate\_d\_minus(D\_minus, idx\_max\_d\_minus)\;
    calculate\_d\_plus(D\_plus, idx\_max\_d\_plus)\;
    
    \eIf{len(results) == 0}{
        results = [max(D\_plus[idx\_max\_d\_plus], D\_minus[idx\_max\_d\_minus]), data.copy()]\;
    }{
        ks = max(D\_plus[idx\_max\_d\_plus], D\_minus[idx\_max\_d\_minus])
        \If{ks $<$ results[0]}{
            results = [ks, data.copy()]\;
        }
    }
    \eIf{D\_plus[idx\_max\_d\_plus] $>$ D\_minus[idx\_max\_d\_minus]}{
        delete data[idx\_max\_d\_plus]\;
    }{
        delete data[idx\_max\_d\_minus]\;
    }
    ++i\;
 }
 \Return results\;
 \caption{Kolmogorov outlier correction}
\end{algorithm}

%% file: discussion.tex
\section{Results and Discussion}

To illustrate some of the DI2 properties, we considered two published datasets: 1) the \textit{breast-tissue} \textit{dataset} \cite{jossinet1996variability}, containing electrical impedance measurements in samples of freshly excised tissue from the breast, and 2) the \textit{yeast} \textit{dataset} \cite{horton1996probabilistic}, containing molecular statistics variables. Both of these are available at the UCI Machine Learning repository \cite{asuncion2007uci}.

This section first discusses results on \textit{breast-tissue dataset}, DI2 is executed with $\tilde{\chi}^2$ as the main statistical test, with and without Kolmogorov outlier removal, with single column discretization, and 5 categories per variable outputted. We will then present and discuss the fitting of the distribution with and without Kolmogorov outlier removal and the interval borders created compared with equal-frequency and equal-width.

We then consider \textit{yeast dataset}, DI2 is executed with $\tilde{\chi}^2$ as the main statistical test, with and without Kolmogorov outlier removal, with single column and whole dataset discretization, and 5 categories per variable outputted, We will then present and discuss the distribution of the values by category with and without border values and compare the different executions of DI2 with themselves and with K-means discretization category distribution.

Finally, still considering the \textit{yeast dataset}, we will present the execution of multiple algorithms and the accuracy achieved with different DI2 discretization configurations and the other aforementioned discretization techniques.

\subsection{\textit{breast-tissue dataset}}

\begin{figure*}[!ht]
    \centering
    \subfigure[\small Q-Q plot of empirical distribution (blue dots) against the fitted \textit{recipinvgauss} distribution (red line).]{\includegraphics[scale=0.3]{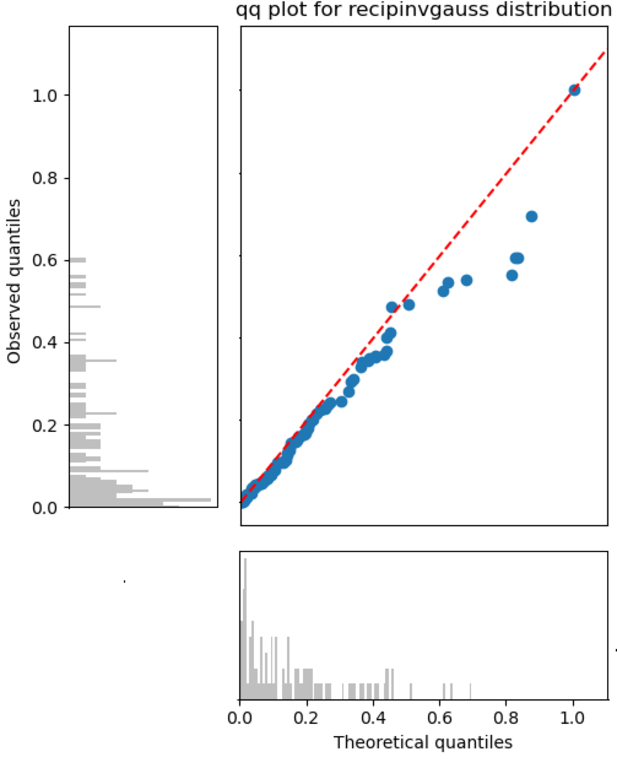}\label{fig:un_optimization}}
    \hfill
    \subfigure[\small Q-Q plot of empirical distribution (blue dots) against the fitted \textit{chi2} distribution (red line). ]{\includegraphics[scale=0.3]{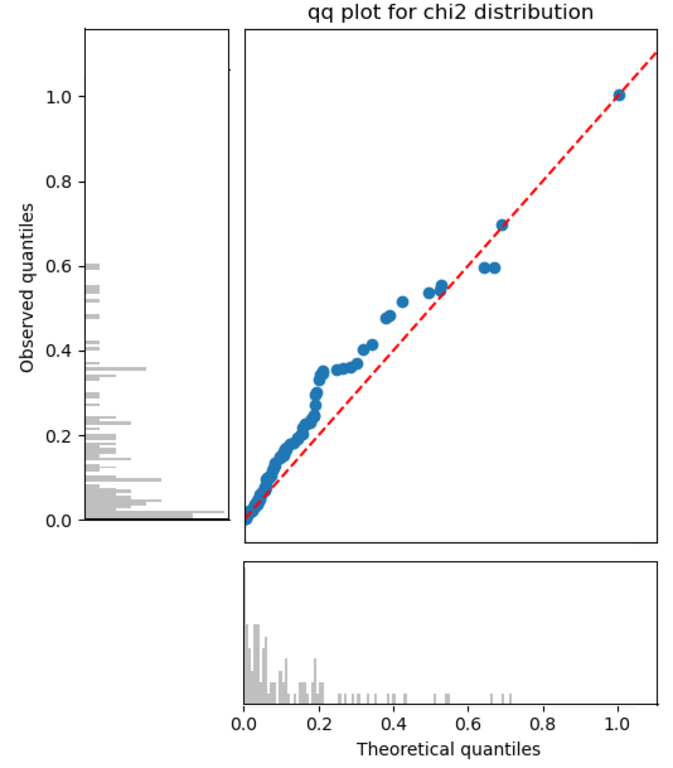}\label{fig:optimization}}
    \hfill
    \subfigure[\small Empirical distribution (gray bins) and corresponding cut-off points using equal-width, equal-frequency and D2I statistical fitting with and without Kolmogorov-Smirnov correction. Red and yellow lines correspond to category and border boundaries.]{\includegraphics[scale=0.4]{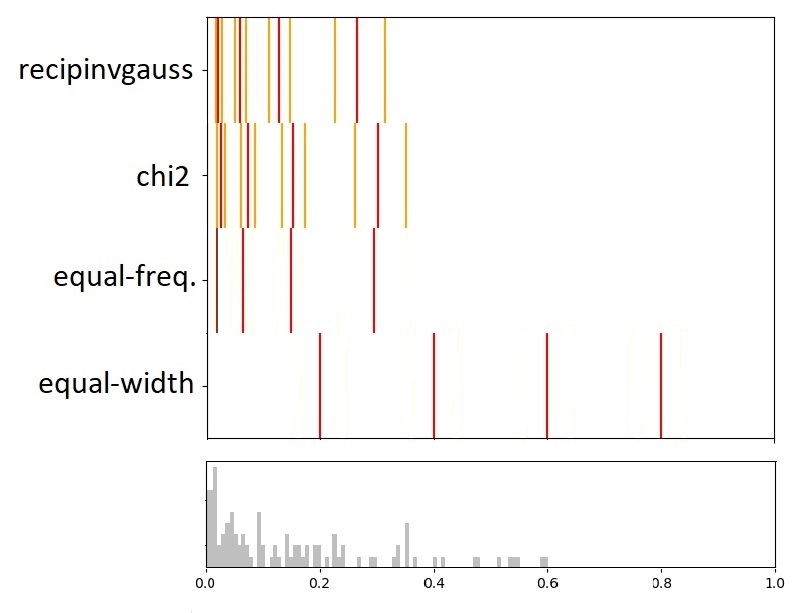}\label{fig:discretization}}
    \label{discretizations}
    \caption{Distribution matching of DA variable from breast-tissue againt two statistical distributions (\textit{recipinvgauss} in V.a and \textit{chi2} in V.b, as well as the corresponding discretization boundaries and border values (V.c).}
    %\vskip -0.5cm
\end{figure*}

The \textit{breast-tissue} dataset contains 106 data instances and 9 continuous variables (I0, PA500, HFS, DA, AREA, A/DA, MAX IP, DR, P). The gathered results show the decisions placed by DI2 in the absence and presence of Kolmogorov-Smirnov optimization. %For this analysis, we considered a min-max normalization for all variables, a desirable number of 5 categories per variable, and $\tilde{\chi}^2$ as the primary statistical test. 

Table \ref{fitting} shows the best fitting distribution for each continuous variable of the dataset. %You can detect 3 types of results in this table. 
Variables "I0", "PA500", "A/DA", "DR", and "P" remained unchanged with a removal of up to 5\% of outlier points. Variables "HFS" and "Area" produced better results in the $\tilde{\chi}^2$ test with the removal of outliers solidifying the distribution choice. Finally, the fitting choice changed for variables "DA" and "Max IP" under the $\tilde{\chi}^2$ test, revealing a more solid choice from the analysis of the residuals.

\begin{table}[!ht]
    \caption{\small Best fitting distributions for each continuous variable, without and with Kolmogorov-Smirnov correction. Both $\tilde{\chi}^2$ (primary) and KS statistics are shown.} %but with a focus on the $\tilde{\chi}^2$.}
    \centering
    \scriptsize
    \begin{tabular}{p{1cm}|p{1.3cm}p{0.9cm}p{0.8cm}p{0.9cm}|p{1.3cm}p{0.9cm}p{0.8cm}p{0.9cm}p{1cm}p{1cm}}
    \toprule
           Variables & Without opt.               & $\tilde{\chi}^2$ statistic & p-value \textless 0.05            & D statistic   & With opt.             & $\tilde{\chi}^2$ statistic  & p-value \textless 0.05         & D statistic           \\ \midrule
    I0     & \textbf{alpha}         & \textbf{8.8} & True & {0.12}   & \textbf{alpha}      & \textbf{8.8} &  True & {0.11}  \\
    PA500  & \textbf{exponnorm}     & \textbf{2.98} & True & {0.07}  & \textbf{exponnorm}  & \textbf{2.98} & True & {0.07} \\
    HFS    & \textbf{foldcauchy}    & \textbf{2.25} & True & {0.07}   & \textbf{foldcauchy} & \textbf{1.57} & True & {0.07}  \\
    DA     & \textbf{recipinvgauss} & \textbf{1.6} &  True & {0.06}   & \textbf{chi2}       & \textbf{1.01} & True & {0.06} \\
    Area   & \textbf{frechet\_r}    & \textbf{0.5} &  True & {0.07}   & \textbf{frechet\_r} & \textbf{0.25} & True & {0.05}  \\
    A/DA   & \textbf{mielke}        & \textbf{1.17} & True & {0.06}   & \textbf{mielke}     & \textbf{1.17} & True & {0.05} \\
    Max IP & \textbf{johnsonsu}     & \textbf{4.72} & True & {0.05}   & \textbf{alpha}      & \textbf{1.09} & True & {0.07}  \\
    DR     & \textbf{johnsonsb}     & \textbf{1.2} & True  & {0.05}   & \textbf{johnsonsb}  & \textbf{1.2} & True & {0.05}  \\
    P      & \textbf{genextreme}    & \textbf{5.13} & True & {0.09}    & \textbf{genextreme} & \textbf{5.13} & True & {0.09}  \\ 
    \end{tabular}
    \label{fitting}
\end{table}

Considering variable "DA", %as an example. %and analyse in more detail. Using 
Figures V.a and V.b %\ref{fig:un_optimization} and \ref{fig:optimization} 
show its Q-Q (quantile-quantile) plot, offering a view on the adequacy of the statistical fitting.  %\citep{wilk1968probability} shows how its observations fit the theoretical continuous distribution. 
In this context, we depict histograms for the observed data with 100 bins (blue dots) and the best theoretical distribution picked without and with Kolmogorov-Smirnov correction (red line). A moderate improvement from Figure V.a to V.b can be detected, with the observed quantiles (blue dots) being closer to the theoretical continuous quantiles (red line). 

After the fitting stage, cut-off points are calculated to produce the final categories. % using the percent point function of the theoretical continuous distribution. 
Figure V.c 
compares different discretization options: %histogram of the observed distribution, %the categories  
%and the border cut off points of each category are marked as yellow lines. 
%We can observe where the equal-width and 
equal-frequency, equal-width, and the two best fitting theoretical continuous distributions (without and with Kolmogorov-Smirnov optimization). Cut-off points are marked as red lines, and the border cut-off points in yellow. 
This analysis shows how critical discretization can be for determining the inclusion or exclusion of high density bins. The ability of DI2 to assign multiple items using borders can thus be explored by symbolic approaches to mitigate vulnerabilities inherent to the discretization process \cite{ushakovrecovering, chmielewski1994global}.

\subsection{\textit{yeast dataset}}

The \textit{yeast dataset} contains 1484 data instances and 10 variables, including the sample identification, class, and 8 molecular statistics variables (mcg, gvh, alm, mit, erl, pox, vac, nuc). %The aforementioned variables represent: 1) McGeoch's method  for signal sequence recognition; 2) von Heijne's method for signal sequence recognition; 3) Score of the ALOM membrane spanning region prediction program; 4) Score of discriminant analysis of the amino acid content of the N-terminal region (20 residues long) of mitochondrial and non-mitochondrial proteins; 5) Presence of "HDEL" substring (thought to act as a signal for retention in the endoplasmic reticulum lumen); 6) Peroxisomal targeting signal in the C-terminus; 7)  Score of discriminant analysis of the amino acid content of vacuolar and extracellular proteins; and 8) Score of discriminant analysis of nuclear localization signals of nuclear and non-nuclear proteins.
In the \textit{breast-tissue dataset} we compared DI2 with two other unsupervised discretization, equal-frequency and equal-width. With this dataset we will compare DI2 with k-means. 

Table \ref{fitting2} displays the results of the statistical tests produced by DI2 when applied to each variable independently and, the last row of the table, when applied to the whole dataset together. Let's use variable "mit" as an example for this dataset. Figure VI.a displays the distribution of values in the variable "mit" before outlier removal (brown and blue area of histogram) and after outlier removal (brown area of histogram). 

%não sei como referenciar https://scikit-learn.org/stable/modules/generated/sklearn.preprocessing.KBinsDiscretizer.html

Figures VI.b and VI.c compare the distribution of the categories of both discretization techniques (DI2 with outlier removal along different discretizations and k-means discretization), and also assess the impact of outlier removal had in categorizing in different executions of DI2 and k-means. Figure VII presents the border values under different DI2 discretization settings.

\begin{figure*}[!htp]
    \centering
    \subfigure[\small Variable "mit" distribution before Kolmogorov-Smirnov outlier removal (brown and blue area) after Kolmogorov-Smirnov outlier removal (brown area) .]{\includegraphics[width=0.45\textwidth]{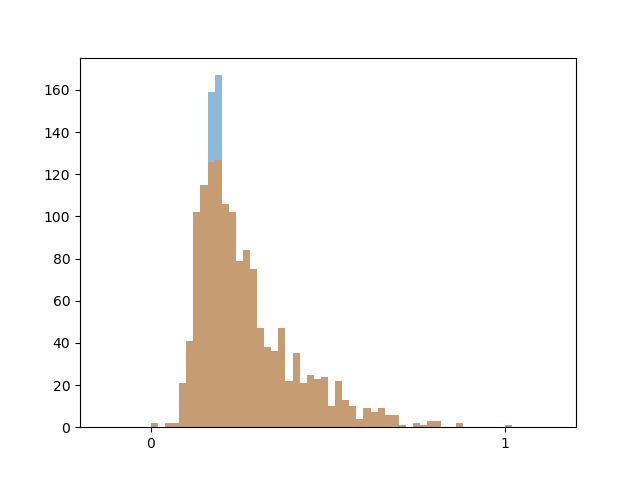}\label{fig:mit_DI2_single_with_nobordervalues}}
    \hfill
    \centering
    \subfigure[\small Variable "mit" categories distribution after DI2 single column discretization, Kolmogorov-Smirnov outlier removal and no border values (blue columns) and after k-means discretization (pink columns). ]{\includegraphics[width=0.45\textwidth]{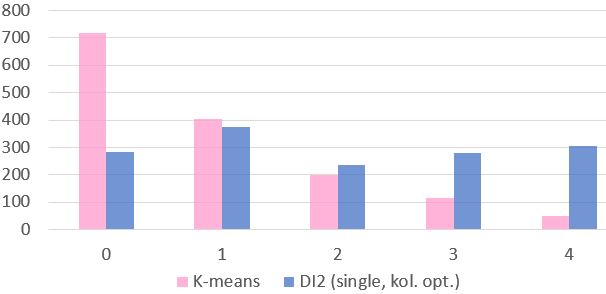}\label{fig:mit_DI2_single_without_nobordervalues}}
    \centering
    \subfigure[\small Variable "mit" categories distribution after DI2 discretization with different settings without border values. Single column discretization without Kolmogorov-Smirnov outlier removal (dark blue columns), whole dataset discretization with Kolmogorov-Smirnov outlier removal (light purple columns), whole discretization without Kolmogorov-Smirnov outlier removal (dark purple columns). ]{\includegraphics[width=0.9\textwidth]{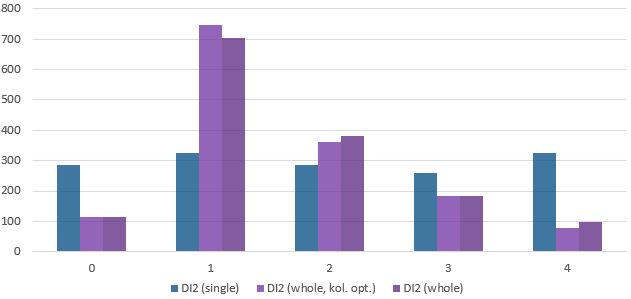}\label{fig:mit_DI2_single_with_bordervalues}}
    \caption{Variable "mit" distribution, IV.a., categories distribution after DI2, IV.b. and IV.c., and after k-means discretization, IV.b.}
    \label{discretizations1}
\end{figure*}

\begin{figure*}[!htp]
    \centering
    \subfigure[\small Categories between 0 and 2.]{\includegraphics[width=0.9\textwidth]{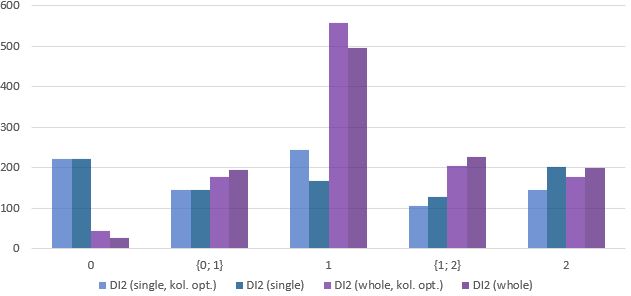}\label{fig:mit_DI2_whole_with_nobordervalues}}
    \hfill
    \subfigure[\small Categories between 2 and 4.]{\includegraphics[width=0.9\textwidth]{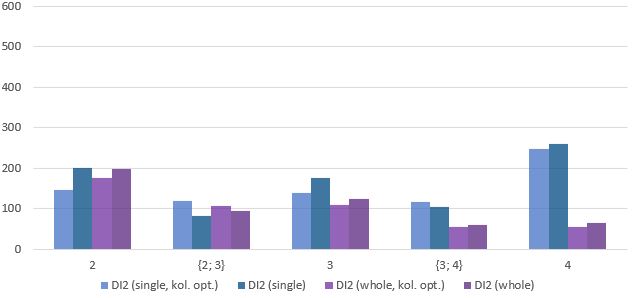}\label{fig:mit_DI2_whole_without_nobordervalues}}
    \caption{Variable "mit" categories distribution after DI2 discretization with different settings with border values. Single column discretization with Kolmogorov-Smirnov outlier removal (light blue columns), single column discretization without Kolmogorov-Smirnov outlier removal (dark blue columns), whole dataset discretization with Kolmogorov-Smirnov outlier removal (light purple columns), whole discretization without Kolmogorov-Smirnov outlier removal (dark purple columns).}
    \label{discretizations2}
\end{figure*}

%Table \ref{fitting} shows the best fitting distribution for each continuous variable of the dataset without and with Kolmogorov-Smirnov outlier removal. Variables "I0", "PA500", "A/DA", "DR", and "P" remained unchanged with a removal of up to 5\% of outlier points. Variables "HFS" and "Area" produced better results in the $\tilde{\chi}^2$ test with the removal of outliers solidifying the distribution choice. Finally, the fitting choice changed for variables "DA" and "Max IP" under the $\tilde{\chi}^2$ test, revealing a more solid choice from the analysis of the residuals.

\begin{table}[!ht]
    \caption{\small Best fitting distributions for each continuous variable, without and with Kolmogorov- Smirnov outlier removal.}
    %but with a focus on the $\tilde{\chi}^2$.}
    \centering
    \scriptsize
    \begin{tabular}{p{1.3cm}|p{1.3cm}p{0.9cm}p{0.8cm}p{0.9cm}|p{1.3cm}p{0.9cm}p{0.8cm}p{0.9cm}p{1cm}p{1cm}}
    \toprule
        Variables & Without opt. & 
        $\tilde{\chi}^2$ statistic  & 
        p-value \textless 0.05 & 
        D statistic  & 
        With opt. & 
        $\tilde{\chi}^2$ statistic &
        p-value \textless 0.05 &
        D statistic \\ \midrule
        mcg     & \textbf{foldcauchy}   & 3.72  & True  & {0.08}    & \textbf{exponnorm}   & 3.18  & True & {0.02} \\
        gvh     & \textbf{genlogistic}  & 3.57  & True & {0.03}    & \textbf{genlogistic}      & 2.02  & True & {0.02} \\
        alm     & \textbf{genlogistic}  & 17.00  & False  & {0.05}    & \textbf{genlogistic} & 12.08  & False & {0.03} \\
        mit     & \textbf{exponnorm}    & 19.23  & False  & {0.05}    & \textbf{exponnorm}   & 6.11  & True & {0.03} \\
        erl     & \textbf{chi2}         & 4.45 $\times 10^{-14}$  & True & {0.99}    & \textbf{chi2}    & 4.23 $\times 10^{-14}$  & True & {0.99} \\
        pox     & \textbf{chi2}         & 4.45 $\times 10^{-14}$  & True & {0.99}    & \textbf{gengamma}    & 4.23 $\times 10^{-14}$  & True & {0.99} \\
        vac     & \textbf{laplace}      & 20.99  & False & {0.08}    & \textbf{pearson3} & 14.18  & False & {1.00} \\
        nuc     & \textbf{exponnorm}    & 1116.63  & False & {0.26}    & \textbf{mielke}  & 795.28  & False & {0.26} \\
        \textbf{all variables} & \textbf{genhalflogistic} & 45.69  & False & {0.25}    & \textbf{genhalflogistic}  & 10.25  & False & {0.21}  \\
    \end{tabular}
    \label{fitting2}
\end{table}

The performed analysis for the \textit{yeast dataset} shows how critical the category border, previously discussed in more detail with the \textit{breast-tissue} dataset, can be, determining the inclusion or exclusion of values. The ability of DI2 to assign multiple items using borders can be explored by symbolic approaches to mitigate vulnerabilities inherent to the discretization process as discussed in the following subsection.

\subsection{Predictive performance}

To test the impact DI2 has when discretizing data we considered the \textit{yeast dataset}, 5 categories per variable, and six supervised classification methods: Naive Bayes \cite{John1995}, Random Forest \cite{Breiman2001}, SMO \cite{platt1998sequential}(Sequential Minimal Optimization), C4.5 \cite{quinlan2014c4}, MLRM \cite{leCessie1992}(Multinomial Logistic Regression Model), and FleBiC \cite{HENRIQUES2021107900}. 

\begin{figure*}[!htp]
    \centering
    \includegraphics[width=0.9\textwidth]{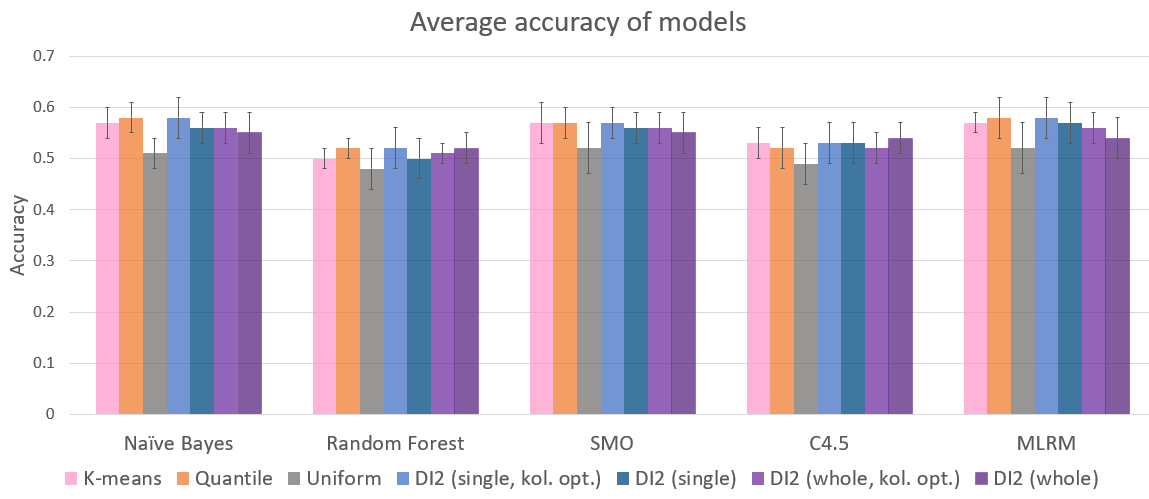}
    \caption{Average accuracy per classifier and discretization method available without border values. From left to right in each group of bars: K-means, Quartile, Uniform, DI2 (single, kol. correction), DI2 (single), DI2 (whole, kol. correction), DI2 (whole).
}
    \label{models_accuracy}
\end{figure*}

In Figure \ref{models_accuracy} the results for the aforementioned models are presented, with the exception of FleBiC which will be discussed later on. Each bar represents the average accuracy achieved with each discretization method, and the small bracket the standard deviation of the accuracy. In each model, DI2, with configurations of single column discretization and outlier removal, matched with the highest accuracy achieved by other discretization methods, and for the C4.5 model, DI2, with configurations of combined column discretization, achieved the highest accuracy compared with other discretization methods. 

\begin{figure*}[!htp]
    \centering
    \subfigure[\small Average accuracy of each FleBiC with different discretization methods.]{\includegraphics[width=0.9\textwidth]{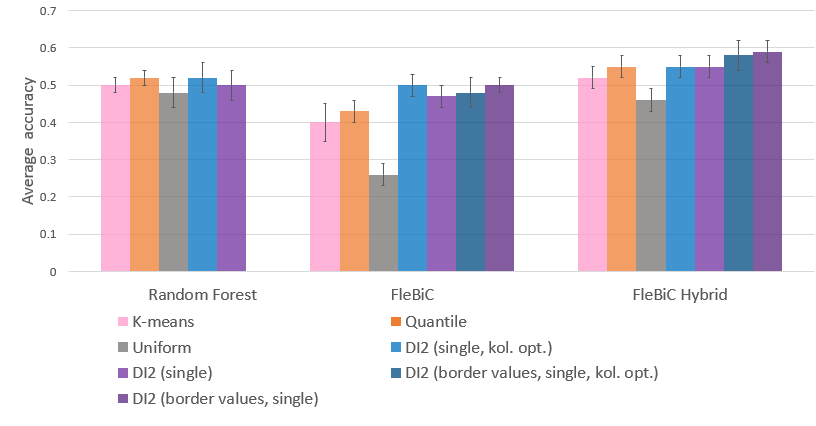}\label{fig:flebic}}
    \centering
    \subfigure[\small Average sensitivity when predicting class NUC of each FleBiC version.]{\includegraphics[width=0.9\textwidth]{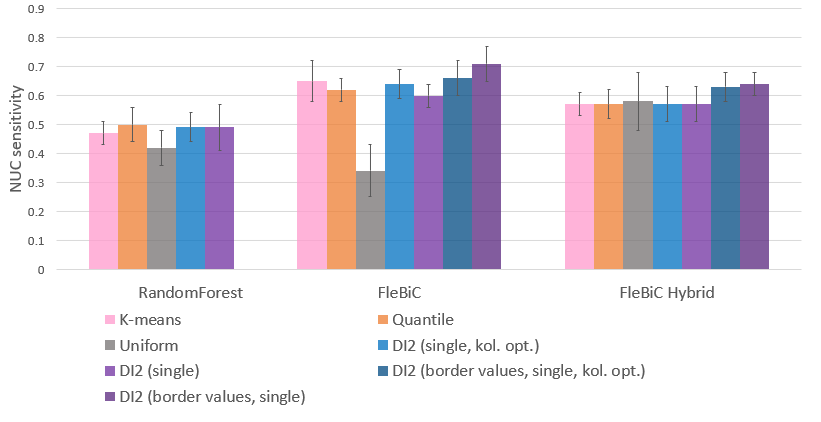}\label{fig:flebic_NUC}}
    \caption{Accuracy when executing different FleBiC versions, and Sensitivity of when predicting class NUC, with multiple discretization methods. From left to right the bars are: K-means, Quartile, Uniform, DI2 (single, kol. correction), DI2 (single), DI2 (border values, single, kol. correction), DI2 (border values, single).}
    \label{FleBiC}
\end{figure*}

In order to fully test out the potential of DI2, we now consider border values. From the aforementioned supervised classification methods only FleBiC is able to test this feature. FleBiC can be executed normally or with Random Forests (weight of decision 50\% FleBiC, 50\% Random Forests), which we will designate as FleBiC Hybrid. In Figure \ref{FleBiC} the results of executing FleBiC and FleBiC Hybrid are presented. In terms of average accuracy, Figure \ref{FleBiC}.a., both FleBiC and FleBiC Hybrid predict with a higher average accuracy when using DI2 method than when using other discretization methods. Within the different settings of DI2, FleBiC Hybrid presents a higher accuracy when the predictive model considers border values. Finally in terms of sensitivity, Figure \ref{FleBiC}b., we can see how considering border values affects the prediction of class NUC, making it possible to break through a ceiling on the prediction of class NUC when other discretization methods couldn't.

%% file: bmc_article.bbl

\begin{thebibliography}{27}
\ifx \bisbn   \undefined \def \bisbn  #1{ISBN #1}\fi
\ifx \binits  \undefined \def \binits#1{#1}\fi
\ifx \bauthor  \undefined \def \bauthor#1{#1}\fi
\ifx \batitle  \undefined \def \batitle#1{#1}\fi
\ifx \bjtitle  \undefined \def \bjtitle#1{#1}\fi
\ifx \bvolume  \undefined \def \bvolume#1{\textbf{#1}}\fi
\ifx \byear  \undefined \def \byear#1{#1}\fi
\ifx \bissue  \undefined \def \bissue#1{#1}\fi
\ifx \bfpage  \undefined \def \bfpage#1{#1}\fi
\ifx \blpage  \undefined \def \blpage #1{#1}\fi
\ifx \burl  \undefined \def \burl#1{\textsf{#1}}\fi
\ifx \doiurl  \undefined \def \doiurl#1{\textsf{#1}}\fi
\ifx \betal  \undefined \def \betal{\textit{et al.}}\fi
\ifx \binstitute  \undefined \def \binstitute#1{#1}\fi
\ifx \binstitutionaled  \undefined \def \binstitutionaled#1{#1}\fi
\ifx \bctitle  \undefined \def \bctitle#1{#1}\fi
\ifx \beditor  \undefined \def \beditor#1{#1}\fi
\ifx \bpublisher  \undefined \def \bpublisher#1{#1}\fi
\ifx \bbtitle  \undefined \def \bbtitle#1{#1}\fi
\ifx \bedition  \undefined \def \bedition#1{#1}\fi
\ifx \bseriesno  \undefined \def \bseriesno#1{#1}\fi
\ifx \blocation  \undefined \def \blocation#1{#1}\fi
\ifx \bsertitle  \undefined \def \bsertitle#1{#1}\fi
\ifx \bsnm \undefined \def \bsnm#1{#1}\fi
\ifx \bsuffix \undefined \def \bsuffix#1{#1}\fi
\ifx \bparticle \undefined \def \bparticle#1{#1}\fi
\ifx \barticle \undefined \def \barticle#1{#1}\fi
\ifx \bconfdate \undefined \def \bconfdate #1{#1}\fi
\ifx \botherref \undefined \def \botherref #1{#1}\fi
\ifx \url \undefined \def \url#1{\textsf{#1}}\fi
\ifx \bchapter \undefined \def \bchapter#1{#1}\fi
\ifx \bbook \undefined \def \bbook#1{#1}\fi
\ifx \bcomment \undefined \def \bcomment#1{#1}\fi
\ifx \oauthor \undefined \def \oauthor#1{#1}\fi
\ifx \citeauthoryear \undefined \def \citeauthoryear#1{#1}\fi
\ifx \endbibitem  \undefined \def \endbibitem {}\fi
\ifx \bconflocation  \undefined \def \bconflocation#1{#1}\fi
\ifx \arxivurl  \undefined \def \arxivurl#1{\textsf{#1}}\fi
\csname PreBibitemsHook\endcsname

%%% 1
\bibitem{altman2014categorizing}
\begin{botherref}
\oauthor{\bsnm{Altman}, \binits{D.G.}}:
Categorizing continuous variables.
Wiley StatsRef: Statistics Reference Online
(2014)
\end{botherref}
\endbibitem

%%% 2
\bibitem{bennette2012against}
\begin{barticle}
\bauthor{\bsnm{Bennette}, \binits{C.}},
\bauthor{\bsnm{Vickers}, \binits{A.}}:
\batitle{Against quantiles: categorization of continuous variables in
  epidemiologic research, and its discontents}.
\bjtitle{BMC medical research methodology}
\bvolume{12}(\bissue{1}),
\bfpage{21}
(\byear{2012})
\end{barticle}
\endbibitem

%%% 3
\bibitem{liao2002appropriate}
\begin{barticle}
\bauthor{\bsnm{Liao}, \binits{S.-C.}},
\bauthor{\bsnm{Lee}, \binits{I.-N.}}:
\batitle{Appropriate medical data categorization for data mining classification
  techniques}.
\bjtitle{Medical informatics and the Internet in medicine}
\bvolume{27}(\bissue{1}),
\bfpage{59}--\blpage{67}
(\byear{2002})
\end{barticle}
\endbibitem

%%% 4
\bibitem{henriques2014bicpam}
\begin{barticle}
\bauthor{\bsnm{Henriques}, \binits{R.}},
\bauthor{\bsnm{Madeira}, \binits{S.C.}}:
\batitle{Bicpam: Pattern-based biclustering for biomedical data analysis}.
\bjtitle{AMB}
\bvolume{9}(\bissue{1}),
\bfpage{27}
(\byear{2014})
\end{barticle}
\endbibitem

%%% 5
\bibitem{chen2016xgboost}
\begin{bchapter}
\bauthor{\bsnm{Chen}, \binits{T.}},
\bauthor{\bsnm{Guestrin}, \binits{C.}}:
\bctitle{Xgboost: A scalable tree boosting system}.
In: \bbtitle{Proceedings of the 22nd Acm Sigkdd International Conference on
  Knowledge Discovery and Data Mining},
pp. \bfpage{785}--\blpage{794}
(\byear{2016})
\end{bchapter}
\endbibitem

%%% 6
\bibitem{okada2007exhaustive}
\begin{barticle}
\bauthor{\bsnm{Okada}, \binits{Y.}},
\bauthor{\bsnm{Okubo}, \binits{K.}},
\bauthor{\bsnm{Horton}, \binits{P.}},
\bauthor{\bsnm{Fujibuchi}, \binits{W.}}:
\batitle{Exhaustive search method of gene expression modules and its
  application to human tissue data}.
\bjtitle{IAENG international journal of computer science}
\bvolume{34}(\bissue{1}),
\bfpage{119126}
(\byear{2007})
\end{barticle}
\endbibitem

%%% 7
\bibitem{zhang2017hierarchical}
\begin{barticle}
\bauthor{\bsnm{Zhang}, \binits{L.}},
\bauthor{\bsnm{Shah}, \binits{S.K.}},
\bauthor{\bsnm{Kakadiaris}, \binits{I.A.}}:
\batitle{Hierarchical multi-label classification using fully associative
  ensemble learning}.
\bjtitle{Pattern Recognition}
\bvolume{70},
\bfpage{89}--\blpage{103}
(\byear{2017})
\end{barticle}
\endbibitem

%%% 8
\bibitem{wang2017multi}
\begin{botherref}
\oauthor{\bsnm{Wang}, \binits{T.}}:
Multi-value rule sets.
arXiv preprint arXiv:1710.05257
(2017)
\end{botherref}
\endbibitem

%%% 9
\bibitem{wang2018multi}
\begin{bchapter}
\bauthor{\bsnm{Wang}, \binits{T.}}:
\bctitle{Multi-value rule sets for interpretable classification with
  feature-efficient representations}.
In: \bbtitle{Proceedings of the 32nd International Conference on Neural
  Information Processing Systems},
pp. \bfpage{10858}--\blpage{10868}
(\byear{2018})
\end{bchapter}
\endbibitem

%%% 10
\bibitem{garcia2012survey}
\begin{barticle}
\bauthor{\bsnm{Garcia}, \binits{S.}},
\bauthor{\bsnm{Luengo}, \binits{J.}},
\bauthor{\bsnm{S{\'a}ez}, \binits{J.A.}},
\bauthor{\bsnm{Lopez}, \binits{V.}},
\bauthor{\bsnm{Herrera}, \binits{F.}}:
\batitle{A survey of discretization techniques: Taxonomy and empirical analysis
  in supervised learning}.
\bjtitle{IEEE Transactions on Knowledge and Data Engineering}
\bvolume{25}(\bissue{4}),
\bfpage{734}--\blpage{750}
(\byear{2012})
\end{barticle}
\endbibitem

%%% 11
\bibitem{yang2009discretization}
\begin{barticle}
\bauthor{\bsnm{Yang}, \binits{Y.}},
\bauthor{\bsnm{Webb}, \binits{G.I.}}:
\batitle{Discretization for naive-bayes learning: managing discretization bias
  and variance}.
\bjtitle{Machine learning}
\bvolume{74}(\bissue{1}),
\bfpage{39}--\blpage{74}
(\byear{2009})
\end{barticle}
\endbibitem

%%% 12
\bibitem{tou1974pattern}
\begin{botherref}
\oauthor{\bsnm{Tou}, \binits{J.T.}},
\oauthor{\bsnm{Gonzalez}, \binits{R.C.}}:
Pattern recognition principles
(1974)
\end{botherref}
\endbibitem

%%% 13
\bibitem{dodge2006oxford}
\begin{bbook}
\bauthor{\bsnm{Dodge}, \binits{Y.}},
\bauthor{\bsnm{Commenges}, \binits{D.}}:
\bbtitle{The Oxford Dictionary of Statistical Terms}.
\bpublisher{Oxford University Press on Demand}, \blocation{???}
(\byear{2006})
\end{bbook}
\endbibitem

%%% 14
\bibitem{lowry2014concepts}
\begin{botherref}
\oauthor{\bsnm{Lowry}, \binits{R.}}:
Concepts and applications of inferential statistics
(2014)
\end{botherref}
\endbibitem

%%% 15
\bibitem{gonzalez1977efficient}
\begin{barticle}
\bauthor{\bsnm{Gonzalez}, \binits{T.}},
\bauthor{\bsnm{Sahni}, \binits{S.}},
\bauthor{\bsnm{Franta}, \binits{W.R.}}:
\batitle{An efficient algorithm for the kolmogorov-smirnov and lilliefors
  tests}.
\bjtitle{ACM TOMS}
\bvolume{3}(\bissue{1}),
\bfpage{60}--\blpage{64}
(\byear{1977})
\end{barticle}
\endbibitem

%%% 16
\bibitem{virtanen2020scipy}
\begin{barticle}
\bauthor{\bsnm{Virtanen}, \binits{P.}},
\bauthor{\bsnm{Gommers}, \binits{R.}},
\bauthor{\bsnm{Oliphant}, \binits{T.E.}},
\bauthor{\bsnm{Haberland}, \binits{M.}},
\bauthor{\bsnm{Reddy}, \binits{T.}},
\bauthor{\bsnm{Cournapeau}, \binits{D.}},
\bauthor{\bsnm{Burovski}, \binits{E.}},
\bauthor{\bsnm{Peterson}, \binits{P.}},
\bauthor{\bsnm{Weckesser}, \binits{W.}},
\bauthor{\bsnm{Bright}, \binits{J.}}, \betal:
\batitle{Scipy 1.0: fundamental algorithms for scientific computing in python}.
\bjtitle{Nature methods}
\bvolume{17}(\bissue{3}),
\bfpage{261}--\blpage{272}
(\byear{2020})
\end{barticle}
\endbibitem

%%% 17
\bibitem{jossinet1996variability}
\begin{barticle}
\bauthor{\bsnm{Jossinet}, \binits{J.}}:
\batitle{Variability of impedivity in normal and pathological breast tissue}.
\bjtitle{Medical and biological engineering and computing}
\bvolume{34}(\bissue{5}),
\bfpage{346}--\blpage{350}
(\byear{1996})
\end{barticle}
\endbibitem

%%% 18
\bibitem{horton1996probabilistic}
\begin{bchapter}
\bauthor{\bsnm{Horton}, \binits{P.}},
\bauthor{\bsnm{Nakai}, \binits{K.}}:
\bctitle{A probabilistic classification system for predicting the cellular
  localization sites of proteins.}
In: \bbtitle{Ismb},
vol. \bseriesno{4},
pp. \bfpage{109}--\blpage{115}
(\byear{1996})
\end{bchapter}
\endbibitem

%%% 19
\bibitem{asuncion2007uci}
\begin{botherref}
\oauthor{\bsnm{Asuncion}, \binits{A.}},
\oauthor{\bsnm{Newman}, \binits{D.}}:
UCI machine learning repository
(2007)
\end{botherref}
\endbibitem

%%% 20
\bibitem{ushakovrecovering}
\begin{botherref}
\oauthor{\bsnm{Ushakov}, \binits{N.}},
\oauthor{\bsnm{Ushakov}, \binits{V.}}:
Recovering information lost due to discretization.
In: XXXIV. International Seminar on Stability Problems for Stochastic Models,
p. 102
\end{botherref}
\endbibitem

%%% 21
\bibitem{chmielewski1994global}
\begin{bchapter}
\bauthor{\bsnm{Chmielewski}, \binits{M.R.}},
\bauthor{\bsnm{Grzymala-Busse}, \binits{J.W.}}:
\bctitle{Global discretization of continuous attributes as preprocessing for
  machine learning}.
In: \bbtitle{Third International Workshop on Rough Sets and Soft Computing},
pp. \bfpage{294}--\blpage{301}
(\byear{1994})
\end{bchapter}
\endbibitem

%%% 22
\bibitem{John1995}
\begin{bchapter}
\bauthor{\bsnm{John}, \binits{G.H.}},
\bauthor{\bsnm{Langley}, \binits{P.}}:
\bctitle{Estimating continuous distributions in bayesian classifiers}.
In: \bbtitle{Eleventh Conference on Uncertainty in Artificial Intelligence},
pp. \bfpage{338}--\blpage{345}.
\bpublisher{Morgan Kaufmann},
\blocation{San Mateo}
(\byear{1995})
\end{bchapter}
\endbibitem

%%% 23
\bibitem{Breiman2001}
\begin{barticle}
\bauthor{\bsnm{Breiman}, \binits{L.}}:
\batitle{Random forests}.
\bjtitle{Machine Learning}
\bvolume{45}(\bissue{1}),
\bfpage{5}--\blpage{32}
(\byear{2001})
\end{barticle}
\endbibitem

%%% 24
\bibitem{platt1998sequential}
\begin{botherref}
\oauthor{\bsnm{Platt}, \binits{J.}}:
Sequential minimal optimization: A fast algorithm for training support vector
  machines
(1998)
\end{botherref}
\endbibitem

%%% 25
\bibitem{quinlan2014c4}
\begin{bbook}
\bauthor{\bsnm{Quinlan}, \binits{J.R.}}:
\bbtitle{C4. 5: Programs for Machine Learning}.
\bpublisher{Elsevier}, \blocation{???}
(\byear{2014})
\end{bbook}
\endbibitem

%%% 26
\bibitem{leCessie1992}
\begin{barticle}
\bauthor{\bparticle{le} \bsnm{Cessie}, \binits{S.}},
\bauthor{\bparticle{van} \bsnm{Houwelingen}, \binits{J.C.}}:
\batitle{Ridge estimators in logistic regression}.
\bjtitle{Applied Statistics}
\bvolume{41}(\bissue{1}),
\bfpage{191}--\blpage{201}
(\byear{1992})
\end{barticle}
\endbibitem

%%% 27
\bibitem{HENRIQUES2021107900}
\begin{botherref}
\oauthor{\bsnm{Henriques}, \binits{R.}},
\oauthor{\bsnm{Madeira}, \binits{S.C.}}:
Flebic: Learning classifiers from high-dimensional biomedical data using
  discriminative biclusters with non-constant patterns.
Pattern Recognition,
107900
(2021).
doi:\doiurl{10.1016/j.patcog.2021.107900}
\end{botherref}
\endbibitem

\end{thebibliography}

\newcommand{\BMCxmlcomment}[1]{}

\BMCxmlcomment{

<refgrp>

<bibl id="B1">
  <title><p>Categorizing continuous variables</p></title>
  <aug>
    <au><snm>Altman</snm><fnm>DG</fnm></au>
  </aug>
  <source>Wiley StatsRef: Statistics Reference Online</source>
  <publisher>Wiley Online Library</publisher>
  <pubdate>2014</pubdate>
</bibl>

<bibl id="B2">
  <title><p>Against quantiles: categorization of continuous variables in
  epidemiologic research, and its discontents</p></title>
  <aug>
    <au><snm>Bennette</snm><fnm>C</fnm></au>
    <au><snm>Vickers</snm><fnm>A</fnm></au>
  </aug>
  <source>BMC medical research methodology</source>
  <publisher>Springer</publisher>
  <pubdate>2012</pubdate>
  <volume>12</volume>
  <issue>1</issue>
  <fpage>21</fpage>
</bibl>

<bibl id="B3">
  <title><p>Appropriate medical data categorization for data mining
  classification techniques</p></title>
  <aug>
    <au><snm>Liao</snm><fnm>SC</fnm></au>
    <au><snm>Lee</snm><fnm>IN</fnm></au>
  </aug>
  <source>Medical informatics and the Internet in medicine</source>
  <publisher>Taylor \& Francis</publisher>
  <pubdate>2002</pubdate>
  <volume>27</volume>
  <issue>1</issue>
  <fpage>59</fpage>
  <lpage>-67</lpage>
</bibl>

<bibl id="B4">
  <title><p>BicPAM: Pattern-based biclustering for biomedical data
  analysis</p></title>
  <aug>
    <au><snm>Henriques</snm><fnm>R</fnm></au>
    <au><snm>Madeira</snm><fnm>SC</fnm></au>
  </aug>
  <source>AMB</source>
  <publisher>Springer</publisher>
  <pubdate>2014</pubdate>
  <volume>9</volume>
  <issue>1</issue>
  <fpage>27</fpage>
</bibl>

<bibl id="B5">
  <title><p>Xgboost: A scalable tree boosting system</p></title>
  <aug>
    <au><snm>Chen</snm><fnm>T</fnm></au>
    <au><snm>Guestrin</snm><fnm>C</fnm></au>
  </aug>
  <source>Proceedings of the 22nd acm sigkdd international conference on
  knowledge discovery and data mining</source>
  <pubdate>2016</pubdate>
  <fpage>785</fpage>
  <lpage>-794</lpage>
</bibl>

<bibl id="B6">
  <title><p>Exhaustive search method of gene expression modules and its
  application to human tissue data</p></title>
  <aug>
    <au><snm>Okada</snm><fnm>Y</fnm></au>
    <au><snm>Okubo</snm><fnm>K</fnm></au>
    <au><snm>Horton</snm><fnm>P</fnm></au>
    <au><snm>Fujibuchi</snm><fnm>W</fnm></au>
  </aug>
  <source>IAENG international journal of computer science</source>
  <pubdate>2007</pubdate>
  <volume>34</volume>
  <issue>1</issue>
  <fpage>119126</fpage>
</bibl>

<bibl id="B7">
  <title><p>Hierarchical multi-label classification using fully associative
  ensemble learning</p></title>
  <aug>
    <au><snm>Zhang</snm><fnm>L</fnm></au>
    <au><snm>Shah</snm><fnm>SK</fnm></au>
    <au><snm>Kakadiaris</snm><fnm>IA</fnm></au>
  </aug>
  <source>Pattern Recognition</source>
  <publisher>Elsevier</publisher>
  <pubdate>2017</pubdate>
  <volume>70</volume>
  <fpage>89</fpage>
  <lpage>-103</lpage>
</bibl>

<bibl id="B8">
  <title><p>Multi-value rule sets</p></title>
  <aug>
    <au><snm>Wang</snm><fnm>T</fnm></au>
  </aug>
  <source>arXiv preprint arXiv:1710.05257</source>
  <pubdate>2017</pubdate>
</bibl>

<bibl id="B9">
  <title><p>Multi-value rule sets for interpretable classification with
  feature-efficient representations</p></title>
  <aug>
    <au><snm>Wang</snm><fnm>T</fnm></au>
  </aug>
  <source>Proceedings of the 32nd International Conference on Neural
  Information Processing Systems</source>
  <pubdate>2018</pubdate>
  <fpage>10858</fpage>
  <lpage>-10868</lpage>
</bibl>

<bibl id="B10">
  <title><p>A survey of discretization techniques: Taxonomy and empirical
  analysis in supervised learning</p></title>
  <aug>
    <au><snm>Garcia</snm><fnm>S</fnm></au>
    <au><snm>Luengo</snm><fnm>J</fnm></au>
    <au><snm>S{\'a}ez</snm><fnm>JA</fnm></au>
    <au><snm>Lopez</snm><fnm>V</fnm></au>
    <au><snm>Herrera</snm><fnm>F</fnm></au>
  </aug>
  <source>IEEE Transactions on Knowledge and Data Engineering</source>
  <publisher>IEEE</publisher>
  <pubdate>2012</pubdate>
  <volume>25</volume>
  <issue>4</issue>
  <fpage>734</fpage>
  <lpage>-750</lpage>
</bibl>

<bibl id="B11">
  <title><p>Discretization for naive-Bayes learning: managing discretization
  bias and variance</p></title>
  <aug>
    <au><snm>Yang</snm><fnm>Y</fnm></au>
    <au><snm>Webb</snm><fnm>GI</fnm></au>
  </aug>
  <source>Machine learning</source>
  <publisher>Springer</publisher>
  <pubdate>2009</pubdate>
  <volume>74</volume>
  <issue>1</issue>
  <fpage>39</fpage>
  <lpage>-74</lpage>
</bibl>

<bibl id="B12">
  <title><p>Pattern recognition principles</p></title>
  <aug>
    <au><snm>Tou</snm><fnm>JT</fnm></au>
    <au><snm>Gonzalez</snm><fnm>RC</fnm></au>
  </aug>
  <pubdate>1974</pubdate>
</bibl>

<bibl id="B13">
  <title><p>The Oxford dictionary of statistical terms</p></title>
  <aug>
    <au><snm>Dodge</snm><fnm>Y</fnm></au>
    <au><snm>Commenges</snm><fnm>D</fnm></au>
  </aug>
  <publisher>Oxford University Press on Demand</publisher>
  <pubdate>2006</pubdate>
</bibl>

<bibl id="B14">
  <title><p>Concepts and applications of inferential statistics</p></title>
  <aug>
    <au><snm>Lowry</snm><fnm>R</fnm></au>
  </aug>
  <pubdate>2014</pubdate>
</bibl>

<bibl id="B15">
  <title><p>An efficient algorithm for the Kolmogorov-Smirnov and Lilliefors
  tests</p></title>
  <aug>
    <au><snm>Gonzalez</snm><fnm>T</fnm></au>
    <au><snm>Sahni</snm><fnm>S</fnm></au>
    <au><snm>Franta</snm><fnm>WR</fnm></au>
  </aug>
  <source>ACM TOMS</source>
  <publisher>ACM New York, NY, USA</publisher>
  <pubdate>1977</pubdate>
  <volume>3</volume>
  <issue>1</issue>
  <fpage>60</fpage>
  <lpage>-64</lpage>
</bibl>

<bibl id="B16">
  <title><p>SciPy 1.0: fundamental algorithms for scientific computing in
  Python</p></title>
  <aug>
    <au><snm>Virtanen</snm><fnm>P</fnm></au>
    <au><snm>Gommers</snm><fnm>R</fnm></au>
    <au><snm>Oliphant</snm><fnm>TE</fnm></au>
    <au><snm>Haberland</snm><fnm>M</fnm></au>
    <au><snm>Reddy</snm><fnm>T</fnm></au>
    <au><snm>Cournapeau</snm><fnm>D</fnm></au>
    <au><snm>Burovski</snm><fnm>E</fnm></au>
    <au><snm>Peterson</snm><fnm>P</fnm></au>
    <au><snm>Weckesser</snm><fnm>W</fnm></au>
    <au><snm>Bright</snm><fnm>J</fnm></au>
    <au><cnm>others</cnm></au>
  </aug>
  <source>Nature methods</source>
  <publisher>Nature Publishing Group</publisher>
  <pubdate>2020</pubdate>
  <volume>17</volume>
  <issue>3</issue>
  <fpage>261</fpage>
  <lpage>-272</lpage>
</bibl>

<bibl id="B17">
  <title><p>Variability of impedivity in normal and pathological breast
  tissue</p></title>
  <aug>
    <au><snm>Jossinet</snm><fnm>J</fnm></au>
  </aug>
  <source>Medical and biological engineering and computing</source>
  <publisher>Springer</publisher>
  <pubdate>1996</pubdate>
  <volume>34</volume>
  <issue>5</issue>
  <fpage>346</fpage>
  <lpage>-350</lpage>
</bibl>

<bibl id="B18">
  <title><p>A probabilistic classification system for predicting the cellular
  localization sites of proteins.</p></title>
  <aug>
    <au><snm>Horton</snm><fnm>P</fnm></au>
    <au><snm>Nakai</snm><fnm>K</fnm></au>
  </aug>
  <source>Ismb</source>
  <pubdate>1996</pubdate>
  <volume>4</volume>
  <fpage>109</fpage>
  <lpage>-115</lpage>
</bibl>

<bibl id="B19">
  <title><p>UCI machine learning repository</p></title>
  <aug>
    <au><snm>Asuncion</snm><fnm>A</fnm></au>
    <au><snm>Newman</snm><fnm>D</fnm></au>
  </aug>
  <pubdate>2007</pubdate>
</bibl>

<bibl id="B20">
  <title><p>Recovering information lost due to discretization</p></title>
  <aug>
    <au><snm>Ushakov</snm><fnm>NG</fnm></au>
    <au><snm>Ushakov</snm><fnm>VG</fnm></au>
  </aug>
  <source>XXXIV. International Seminar on Stability Problems for Stochastic
  Models</source>
  <fpage>102</fpage>
</bibl>

<bibl id="B21">
  <title><p>Global discretization of continuous attributes as preprocessing for
  machine learning</p></title>
  <aug>
    <au><snm>Chmielewski</snm><fnm>MR</fnm></au>
    <au><snm>Grzymala Busse</snm><fnm>JW</fnm></au>
  </aug>
  <source>Third international workshop on rough sets and soft
  computing</source>
  <pubdate>1994</pubdate>
  <fpage>294</fpage>
  <lpage>-301</lpage>
</bibl>

<bibl id="B22">
  <title><p>Estimating Continuous Distributions in Bayesian
  Classifiers</p></title>
  <aug>
    <au><snm>John</snm><fnm>GH</fnm></au>
    <au><snm>Langley</snm><fnm>P</fnm></au>
  </aug>
  <source>Eleventh Conference on Uncertainty in Artificial
  Intelligence</source>
  <publisher>San Mateo: Morgan Kaufmann</publisher>
  <pubdate>1995</pubdate>
  <fpage>338</fpage>
  <lpage>345</lpage>
</bibl>

<bibl id="B23">
  <title><p>Random Forests</p></title>
  <aug>
    <au><snm>Breiman</snm><fnm>L</fnm></au>
  </aug>
  <source>Machine Learning</source>
  <pubdate>2001</pubdate>
  <volume>45</volume>
  <issue>1</issue>
  <fpage>5</fpage>
  <lpage>32</lpage>
</bibl>

<bibl id="B24">
  <title><p>Sequential minimal optimization: A fast algorithm for training
  support vector machines</p></title>
  <aug>
    <au><snm>Platt</snm><fnm>J</fnm></au>
  </aug>
  <pubdate>1998</pubdate>
</bibl>

<bibl id="B25">
  <title><p>C4. 5: programs for machine learning</p></title>
  <aug>
    <au><snm>Quinlan</snm><fnm>JR</fnm></au>
  </aug>
  <publisher>Elsevier</publisher>
  <pubdate>2014</pubdate>
</bibl>

<bibl id="B26">
  <title><p>Ridge Estimators in Logistic Regression</p></title>
  <aug>
    <au><snm>Cessie</snm><fnm>S.</fnm></au>
    <au><snm>Houwelingen</snm><fnm>J.C.</fnm></au>
  </aug>
  <source>Applied Statistics</source>
  <pubdate>1992</pubdate>
  <volume>41</volume>
  <issue>1</issue>
  <fpage>191</fpage>
  <lpage>201</lpage>
</bibl>

<bibl id="B27">
  <title><p>FleBiC: Learning classifiers from high-dimensional biomedical data
  using discriminative biclusters with non-constant patterns</p></title>
  <aug>
    <au><snm>Henriques</snm><fnm>R</fnm></au>
    <au><snm>Madeira</snm><fnm>SC</fnm></au>
  </aug>
  <source>Pattern Recognition</source>
  <pubdate>2021</pubdate>
  <fpage>107900</fpage>
  <url>https://www.sciencedirect.com/science/article/pii/S003132032100087X</url>
</bibl>

</refgrp>
} % end of \BMCxmlcomment
